\documentclass{article}

\usepackage[preprint]{neurips_2025}

\usepackage[utf8]{inputenc} 
\usepackage[T1]{fontenc}    
\usepackage{hyperref}       
\usepackage{url}            
\usepackage{booktabs}       
\usepackage{amsfonts}       
\usepackage{nicefrac}       
\usepackage{microtype}      
\usepackage{xcolor}         
\usepackage{times}
\usepackage{soul}

\usepackage{graphicx}
\usepackage{amsmath}
\usepackage{amsthm}
\usepackage{algorithm}
\usepackage{subcaption}
\usepackage{algpseudocode}
\usepackage{tabularx}
\usepackage{booktabs} 
\usepackage{adjustbox}
\usepackage{multirow}
\usepackage{xcolor}
\usepackage{cleveref}
\usepackage{algcompatible}

\title{HoloGraph: All-Optical Graph Learning via Light Diffraction}

\author{
  Yingjie Li\thanks{Equal contribution} \\
  Simon Fraser University\\
  \texttt{yingjie\_li@sfu.ca}
  \And
  Shanglin Zhou\footnotemark[1] \\
  University of Connecticut \\
  \texttt{shanglin.zhou@uconn.edu}
  \And
  Caiwen Ding \\
  University of Minnesota, Twin Cities \\
  \texttt{dingc@umn.edu}
  \And
  Cunxi Yu \\
  University of Maryland, College Park \\
  \texttt{cunxiyu@umd.edu}
}

\begin{document}

\maketitle

\begin{abstract}
As a representative of next-generation device/circuit technology beyond CMOS, physics-based neural networks such as Diffractive Optical Neural Networks (DONNs) have demonstrated promising advantages in computational speed and energy efficiency. However, existing DONNs and other physics-based neural networks have mostly focused on exploring their machine intelligence, with limited studies in handling graph-structured tasks. Thus, we introduce HoloGraph, the first monolithic free-space all-optical graph neural network system. It proposes a novel, domain-specific message-passing mechanism with optical skip channels integrated into light propagation for the all-optical graph learning. HoloGraph enables light-speed optical message passing over graph structures with diffractive propagation and phase modulations. Our experimental results with HoloGraph, conducted using standard graph learning datasets Cora-ML and Citeseer, show competitive or even superior classification performance compared to conventional digital graph neural networks. Comprehensive ablation studies demonstrate the effectiveness of the proposed novel architecture and algorithmic methods.
\end{abstract}

\section{Introduction}
\label{sec:intro}

Conventional Von Neumann architecture faces significant challenges in accommodating the computational and memory requirements of deep neural networks (DNNs), resulting in extensive data movement within the memory hierarchy~\cite{peng2021binary}. In response, researchers are actively exploring various Von Neumann architectures to create efficient DNN accelerators. The current research emphasis centers on two key areas: (i) pushing the boundaries of next-generation device/circuit technology beyond the capabilities of complementary metal-oxide-semiconductor (CMOS) technology, and (ii) collaboratively designing algorithms and technologies that enable real-time DNN processing while consuming minimal power across diverse application domains.



Diffractive Optical Neural Networks (DONNs) have gained substantial attention as a noteworthy and prominent alternative in this field. DONNs utilize natural phenomena such as light diffraction and light signal phase modulation to enable all-optical processing at the speed of light. One significant advantage of DONNs is their considerably lower energy consumption for all-optical inference compared to conventional DNNs on digital platforms, as all-optical computation incurs no additional energy costs to maintain the functionality of computation units (phase modulation pixels in layers) once the phase masks have been fabricated or assembled~\cite{lin2018all}.
The existing DONN system excels in processing highly parallelized regular data structures~\cite{yan2022all}; however, it faces challenges in processing latent and graph-structured information since it has no memory storage and computes with passive devices.


Various scientific fields analyze data beyond the underlying Euclidean domain. For example, the analysis of interconnected entities (nodes) through rich relationships (edges) in graph-structured data is crucial in scientific domains such as chemical molecules~\cite{duvenaud2015convolutional} and brain networks~\cite{hu2013matched}. Graph neural networks (GNNs) and recurrent neural networks (RNNs) have emerged as versatile approaches capable of capturing intricate relationships and dependencies among entities within a graph. The message passing mechanism~\cite{gilmer2020message} is the core operation in GNNs, allowing nodes to communicate and aggregate information from their neighboring nodes. This process iteratively updates the node features based on the features of their neighbors and has been successfully applied in many graph-based applications, including social networks, recommendation systems, and drug discovery~\cite{kipf2016semi,velickovic2017graph,gilmer2017neural}.

While the use of optical domain accelerators has emerged as a promising solution for efficient machine learning acceleration, there are very limited studies addressing graph learning workloads in optical AI systems~\cite{zhou2021large,mengu2019analysis,mengu2020scale,li2022physics}. Diffractive Graph Neural Networks (DGNNs)~\cite{yan2022all} proposed the use of all-optical graph representation learning with diffractive photonic computing units (DPUs) and on-chip optical devices. However, the entire optical AI system suffers from limited scalability due to the expensive devices and delicate fabrication processes. 
In contrast, all free-space optical AI systems, such as DONN, have demonstrated their scalability and cost efficiency in real-world prototyping. Unfortunately, a successful all-passive free-space optical graph learning system does not yet exist due to the complex nature of graph data. The challenges lie not only in processing large volumes of data but also in efficiently encoding graph features (attributes) and constructing effective model structures. Recognizing these challenges, our study introduces HoloGraph, an all-passive optical system specifically designed for graph learning. The all-passive optical system offers significant advantages in energy efficiency compared to conventional GNNs, as well as ease of fabrication at scale compared to other optical graph learning systems~\cite{yan2022all}. In this system, the message passing and aggregation between nodes are realized by trainable optical modulation, while optical skip connections are employed to enhance the latent information during the message passing process. The experimental results include analysis and discussion of various neural architecture configurations. HoloGraph demonstrates competitive accuracy results for classification tasks on real-world benchmark datasets compared with conventional GNNs, underscoring the effectiveness of our approach in graph learning applications. 

\section{Background and Related Work}

\subsection{Free-space Diffractive Optical Neural Network}
\begin{figure}
         \centering
         \includegraphics[width=0.5\linewidth]{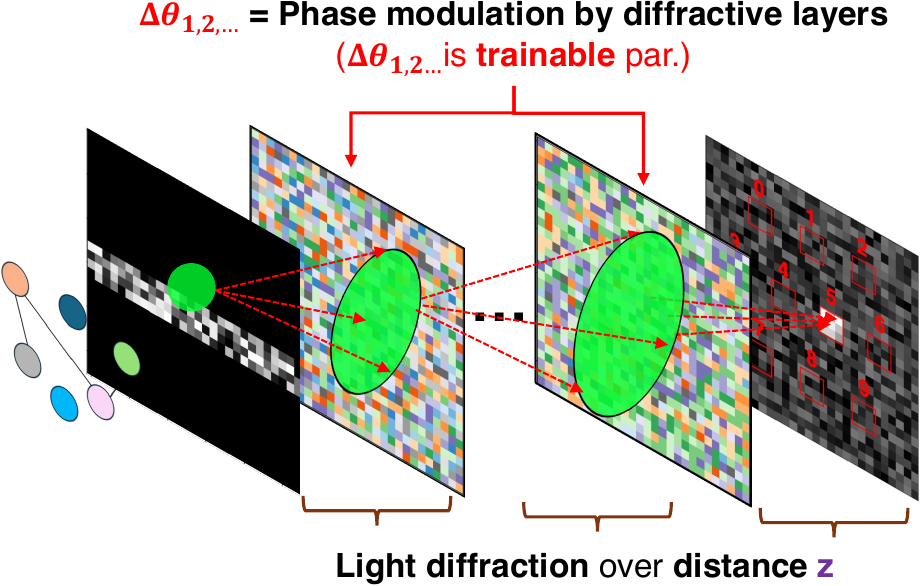}
        \caption{Overview of all-optical multi-layer DONNs system.}
        \label{fig:overview_donn}
\end{figure}

{Diffractive optical neural network (DONN) is a promising research area in optical computing. It features with high system throughput, high computation speed while at low power consumption by all-optical inference with optical passive devices. Compared to conventional neural networks (NNs) on digital platforms, the information carrier changes from electrical signal to light signal in DONN systems. Thus, instead of manipulating electrons between transistors to realize the computation, in DONN systems, the computation is realized by manipulating the information-carried light with its physical features in the free space, i.e., instead of operators such as multiplier, adders in conventional NNs, in DONN systems, the operators are \textit{light diffraction} and \textit{phase modulations}. {As shown in Figure \ref{fig:overview_donn},} the light diffraction happens at the free space between layers over the diffraction distance $z$, and the designed phase modulations $\Delta \theta$, which are the trainable parameters w.r.t the target tasks, are embedded within diffractive layers. By stacking diffractive layers and a detector at the end of the system to capture the light intensity pattern for analysis and predictions, the DONN system deals with computing tasks efficiently. 
DONN systems have demonstrated its efficiency in many applications \cite{lin2018all,yan2022all,liu2023broadband,qian2020performing} and are verified by the practical prototyping \cite{li2022physics,chen2022physics}. However, the studies for its applications in advanced large-scale massage-passing learning, such as graph learning, are limited. }


\paragraph{Free-space Hardware Implementation}



The free-space DONN hardware system consists primarily of three components: the laser source, diffractive layers, and the detector. The diffractive layers are implemented using passive optical devices, where computation is achieved by naturally manipulating the light signal, without requiring additional energy for processing. This results in significant energy efficiency advantages over conventional digital neural networks. Additionally, the free-space DONN system offers remarkable scalability, as input information is represented as an image on the first input layer. The information is encoded in parallel onto the light signal as it propagates through the input layer. Scalability can be easily achieved by increasing the size of the laser beam and expanding the input layer, enabling the encoding of more information.
For example, SLMs can function as diffractive layers in the DONN system with laser wavelength in visible range \cite{chen2022physics}; while for systems with laser wavelength in Terahertz (THz) range, a 3D printed mask with designed thickness at each pixel made with UV-curable resin is used as diffractive layers in THz optical systems \cite{lin2018all}. Compared to integrated photonic processing units \cite{yan2022all,shen2017deep}, the free-space DONN system incurs lower fabrication costs per operator.

\subsection{Graph Neural Network and Message Passing}

Graph neural networks (GNNs) have ascended as the promising methodology for managing non-Euclidean, graph-structured data.
The fundamental operation at the heart of GNNs is the message-passing mechanism, which includes operates by aggregating information from adjacent nodes (and the node itself), to update the representations of the central node with a trainable GNN layer. This process involves a message-passing phase that performs feed-forward inference, succeeded by a readout phase that interprets the nodes' updated states. Given a graph $\mathcal{G(V,E)}$ with the input features ${x_{v} , \forall{v} \in V }$ and the edge $e_{i,j}$, indicating the edge connecting node ${i}$ and ${j}$, the computation within a message-passing GNN with $L$ layers follows a general formula.
\begin{equation}
\begin{split}
\label{equ: message-passing}
    x_i^{l+1} = \sigma (x_i^l, \mathcal{A}_{j \in \mathcal{N}(i)} (\phi (x_i^l, x_j^l, e_{i,j}^l))), ~~~ l = 1, 2, ..., L
\end{split}
\end{equation}
where $x_i^{l+1}$ denotes node $i$'s embedding at the subsequent layer $l+1$, which is is computed by integrating its current embedding $x_i^l$ at layer $l$ with the embedding of its neighboring nodes $j$, represented by $x_j^l$ where $j$ is an element of node $i$'s neighborhood $\mathcal{N}(i)$, as well as the embedding of the edge $e_{i,j}^l$ connecting them. The integration is facilitated through the application of a differentiable message transformation $\phi(\cdot)$, an order-insensitive permutation aggregation function  $\mathcal{A}(\cdot)$  and a differentiable node transformation $\sigma(\cdot)$, which together determine the new state of the node's embedding. 

\paragraph{Optical Message Passing}

\cite{yan2022all} proposed the diffractive GNN (DGNN) for the optical message passing over graph-structured data. It implements diffractive photonic computing units (DPUs) to generate the optical node features. Specifically, the light-encoded initial node features of selected nodes (neighbor nodes and the target node) are input to the integrated DPU to transform the node attributes into optical neural messages, where the strip optical waveguides are deployed to send the transformed output message to the optical couplers to aggregate the optical neural messages from node neighborhoods. Then the DGNN-extracted optical node features are fed into an classifier for node classifications. However, it suffers from the expensive scalability cost regarding the increasing dimensions for neighborhoods and node features. In this work, to avoid the waveguide crossing, they set a 2-dimensional optical message and use the optical Y-coupler for message aggregation. For the node with $n$ neighborhoods, to generate a 2D optical message, the number of required Y-couplers in the system will be proportional to $n^{2}$ and the required DPUs will be proportional to $n$ for a single round message passing.


\section{HoloGraph}
\label{sec:method}

\begin{figure*}
     \centering
     \begin{subfigure}[b]{1\textwidth}
         \centering
         \includegraphics[width=1\linewidth]{ 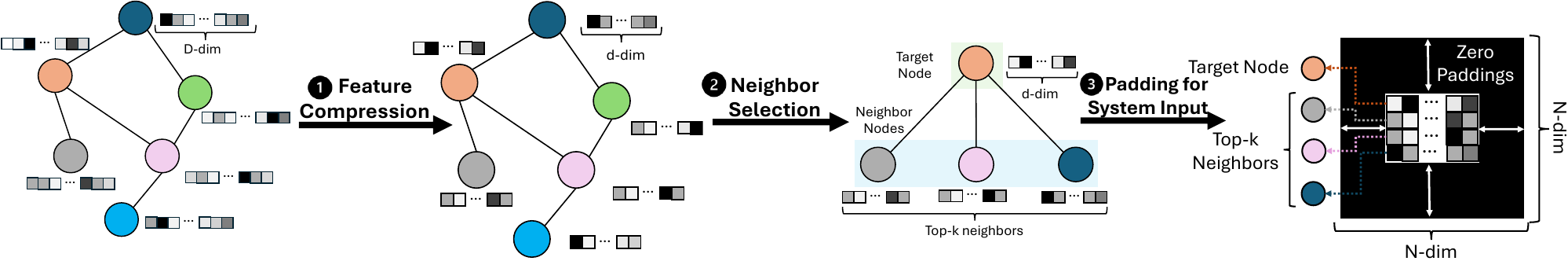}
         \caption{The input preparation of the graph data.}
         \label{fig:input_process}
     \end{subfigure}
     \hfill
    \begin{subfigure}[b]{0.25\textwidth}
         \centering
        \includegraphics[width=0.95\linewidth]{ 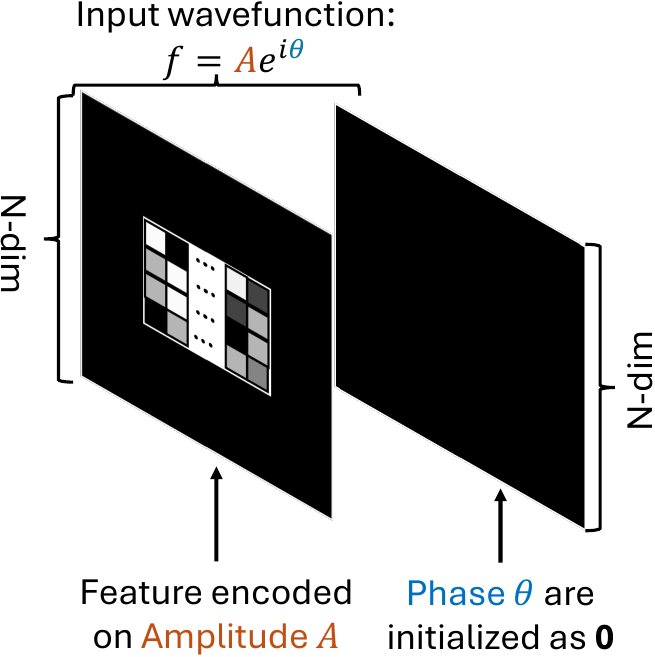}
         \caption{Input encoding with light wavefunction.}
         \label{fig:input_encode}
     \end{subfigure}
     \begin{subfigure}[b]{0.7\textwidth}
         \centering
         \includegraphics[width=1\linewidth]{ 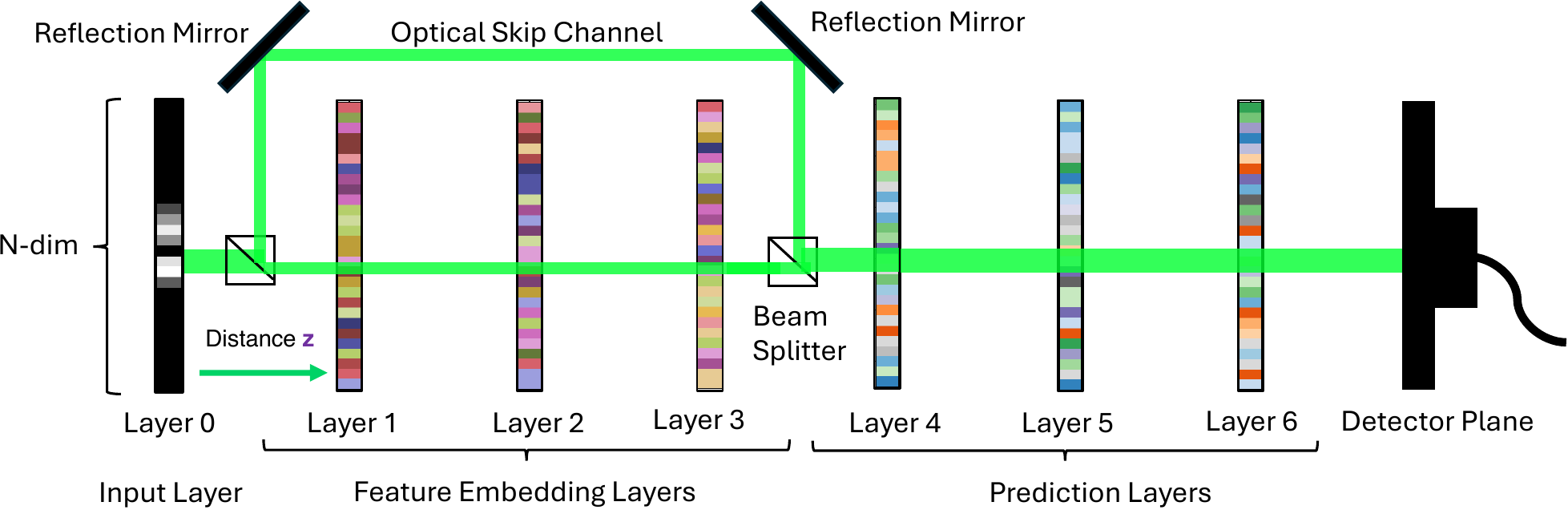}
         \caption{The overview of the network architecture with the optical skip channel connecting the input and the first prediction layer. The optical path is realized by passive optical devices including two beam spliters and two reflection mirrors.}
         \label{fig:sys_overview}
     \end{subfigure}
        \caption{{Illustration of the proposed HoloGraph network with 3 feature embedding layers and 3 prediction layers. The optical skip channel is connecting the input (layer 0) to the first prediction layer (layer 4).}}
        \label{fig:overview}

\end{figure*}

In this section, we will introduce our approach HoloGraph, the first free-space all-optical graph learning framework. Specifically, {as shown in Figure \ref{fig:overview}}, we first pre-process the input data to fit the DONN system. We deploy Principle Component Analysis (PCA)~\cite{abdi2010principal} to preprocess the initial node attributes and reduce their dimensions for each node, and the personalized PageRank algorithm, PPRGo~\cite{bojchevski2020scaling} to directly capture the high-order neighborhood information. Then, we introduce the model structure of HoloGraph, where we adopt residual learning in DONN systems with a optical skip channel, which introduces the previous neural messages into the aggregation with the current neural messages in time sequence. The optical skip channel enables the compensation of information lost during light propagation and smooths the gradient flow during the training process, resulting in more efficient convergence and better accuracy performance.

\subsection{Input Preparation}
\label{subsec:input}

First of all, the graph data is pre-processed to be compatible with the optical neural network. With the system size set to be $N$, we first apply PCA on the original dataset with $D$-dimensional features to reduce the dimension of node attributes to $d$ to fit the DONN system, where $d \leq N$. Then, we pick top-$k$ ($k \leq N$) neighborhoods of the target nodes regarding the graph edgelist with personalized page rank (PPRGo). Then, we can assemble the input matrix, where each neighbor with its corresponding node attributes is a row in the matrix, resulting in the matrix size of $k \times d$. To be compatible with the DONN system, which is a symmetric system, we further pad zeros around the information matrix while centering the information matrix to generate the input with size of $N \times N$. 

{For example, in Figure~\ref{fig:input_process}, all six nodes in the graph originally have $D$ features, representing with $D$-dimensional vectors. First, the features of all the nodes are compressed to $d$-dimensions with PCA. Second, for each node in the graph, it is set as the target node and we select its top-$k$ neighbors according to the set personalized PageRank. For example, for the orange node, we take neighbors including the gray node, the pink node, and the dark blue node to assemble the system input. Third, the features for selected neighbors and the target node are aligned in a matrix. To fit the DONN system, we pad the feature matrix with zeros to the $N \times N$ system input. }

In DONN systems, the information is encoded in complex domain, where richer features can be embedded with the inputs. Specifically, the light signal is described with the wavefunction 
    $f=A\cdot e^{i\theta} \leftrightarrow f=A\cdot\text{cos} \theta + A\cdot i \cdot\text{sin} \theta$,
where $i=\sqrt{-1}$, and $A$ is the \textit{Amplitude} and $\theta$ is the \textit{Phase} of the light signal. With Euler's formula $e^{i\theta} = cos\theta + isin\theta$, the wavefunction is described with a complex number $f=Acos\theta + i Asin\theta$.
Therefore, the input feature embedding is enriched naturally with the complex-valued description in DONN systems. In DONN systems, we encode the initial features on the \textit{Amplitude} while set the \textit{Phase} to be zero as the default setting as shown in Figure~\ref{fig:input_encode}. Additionally, we can optionally encode the personalized PageRank score on the \textit{Phase} by mapping the score in $[0, 1]$ to phase in $[0, \pi/2]$.


\subsection{Network Architectures and Training}
In this section, we will introduce the detailed implementation of HoloGraph including the physical modeling of the DONN system, the implementation of the optical skip channels, and the training with the system.
We set our default DONN system with six diffractive layers and the system size is set to be $200 \times 200$. The illustrated figure of the 6-layer DONN system is shown in Figure \ref{fig:sys_overview}. 

\subsubsection{Forward Function in DONN systems} 
\label{subsubsec:forward}

As illustrated in Section \ref{subsec:input}, the information is encoded with a wavefunction of the light signal described with complex-valued numbers. As shown in Figure \ref{fig:sys_overview}, in the multiple layer stacked DONN system, the information-encoded light signal will be diffracted and phase-modulated iteratively between layers. Finally, the diffraction pattern after light propagation with respect to light intensity distribution is captured at the detector plane for predictions. 

The input information is encoded on the coherent light signal from the laser source, its wavefunction can be expressed as $f_{0}(x_{0}, y_{0}) = Re((x_{0}, y_{0})) + i Im((x_{0}, y_{0}))$, where $(x_{0}, y_{0})$ is the coordinate of the light source at the emission plane. The wavefunction after light diffraction over diffraction distance $z$ to the first diffractive layer is the summation of the outputs at the input plane, 
i.e., 
\begin{equation}
\label{eq:diffraction_time}
    f_{1}(x, y) = \iint f_{0}(x_{0}, y_{0})h(x-x_{0}, y-y_{0}, z)dx_{0}dy_{0}
\end{equation}
where $(x, y)$ is the coordinate on the receive plane. $h$ is the impulse optical response function of free space, which is the mathematical approximation~\cite{tobin1997introduction} for light diffraction. In this work, the forward function is computed with \textit{Fresnel approximation} and the corresponding $h$ is
\begin{equation}
    h(x, y, z) = \frac{e^{ikz}}{i\lambda z}e^{i\frac{k}{2z}(x^2 + y ^2)}
\end{equation}
where $\lambda$ is the wavelength of the light, $k$ is the wave number where $k=\frac{2\pi}{\lambda}$, and $z$ is the free-space diffraction distance.  

Equation \ref{eq:diffraction_time} can be calculated with spectral algorithm, where we employ Fast Fourier Transform (FFT) for fast and differentiable computation. For example, 
    $U_{1}(\alpha, \beta) = U_{0}(\alpha, \beta)H(\alpha, \beta, z)$,
where $U$ and $H$ are the Fourier transformation of $f$ and $h$ respectively. 
The wavefunction $U_{1}(\alpha, \beta)$ after diffraction is transformed to time domain with inverse FFT (iFFT) for phase modulation, i.e.,
    $f_{2}(x, y) = \text{iFFT}(U_{1}(\alpha, \beta)) \times \mathbf{W}_{1}(x, y)$,
where $\mathbf{W}_{1}(x, y)$ is the phase modulation in the first diffractive layer. $f_{2}(x, y)$ is the wavefunction after phase modulation, which is also the input wavefunction for the light diffraction for the second diffractive layer. Thus, the forward function for a multiple-layer-stacked DONN system is computed iteratively.

We wrap one computation round of light diffraction and phase modulation at one diffractive layer as a computation module named \textbf{DiffMSG} (diffractive message passing), i.e.,
    $\text{DiffMSG}(f(x, y), \mathbf{W}) = L(f(x, y), z) \times \mathbf{W}(x, y)$,
where $f(x, y)$ is the input wavefunction.
$\mathbf{W}(x, y)$ is the phase modulation. 
$L(f(x, y), z)$ $=$ $f_{1}(x, y)$ is the wavefunction after light diffraction over distance $z$ in time domain. 
Thus, for a n-layer DONN system, the forward function can be expressed by computing \textbf{DiffMSG} iteratively, i.e., 
\begin{equation}
\label{eq:forward}
\begin{split}
I(f_{0}(x, y), \mathbf{W}) = \text{DiffMSG}(..., \text{DiffMSG}(\text{DiffMSG}(f_{0}(x, y), \mathbf{W}_{1}(x, y)), \mathbf{W}_{2}(x, y), ..., \mathbf{W}_{n}(x, y))...)
\end{split}
\end{equation}
where $\mathbf{W}=\{ \mathbf{W}_{1}, \mathbf{W}_{2}, ..., \mathbf{W}_{n}\}$ is phase modulations for diffractive layers in the DONN system.  

\subsubsection{Optical Skip Channels Implementation}

The existing DONN system~\cite{lin2018all} from Equation~\ref{eq:forward} shows challenges in dealing with graph-structured data without latent information involved in the forward function. Adopted from residual neural networks~\cite{he2016deep}, we implement the optical skip channels into the DONN system to capture the latent information during the forward propagation without memory required in the system. \textbf{Different from the implementation of residual blocks in digital system, the optical skip channel involves light diffraction of the identity information through the skip distance between layers.}
As shown in Figure~\ref{fig:sys_overview}, the first three layers (layer 1 to layer 3) are implemented as feature aggregation layers, which are trained for feature transformation and aggregation of the graph data. Layer 4 to layer 6 are prediction layers where it takes the aggregated feature embedding as input to predict the labels of the target node. The optical skip channel can be implemented between arbitrary layers. For example, in Figure~\ref{fig:sys_overview}, we implement the optical skip channel between the input layer (layer 0) and the first prediction layer (layer 4), where the length of the optical skip channel is 4 times of the diffraction distance $z$ in the DONN system, i.e., the identity information at the first prediction layer (layer 4) is
\begin{equation}
    f^{'}_{0}(x, y) = L(f_{0}(x, y), 4\times{z})
\end{equation}
where $f_{0}(x, y)$ is the input wavefunction at layer 0 and $L$ is the function for light diffraction approximation. Thus, in the forward function, the input wavefunction for layer 4 takes the average of the latent wavefunction $f^{'}_{0}(x, y)$ and the wavefunction from the output of the feature extraction layers, i.e., the forward function with the optical skip channel is 
\begin{equation}
\label{eq:forward_skip}
\begin{split}
    I(f_{0}(x, y), \mathbf{W}) = \text{DiffMSG}(\text{DiffMSG}(\text{DiffMSG} ((\text{DiffMSG}  (\text{DiffMSG}(\text{DiffMSG}(f_{0}(x, y), \\
    \mathbf{W_{1}}), \mathbf{W_{2}}), \mathbf{W_{3}})  {\color{blue}{+ f^{'}_{0}(x, y)})/2}, \mathbf{W_{4}}), \mathbf{W_{5}}), \mathbf{W_{6}}) 
\end{split}
\end{equation}
Note that we can insert more optical skip channels between arbitrary layers while setting the corresponding skip distance in light diffraction for identity information $f^{'}(x, y)$. The optical skip channels can be implemented with passive optical devices including beam splitters and reflection mirrors as shown in Figure~\ref{fig:sys_overview}, which requires no extra energy cost.

\subsubsection{System Training}

The final diffraction pattern with respect to the light intensity \( I \) in Equation \ref{eq:forward_skip} is captured at the detector plane. For classification tasks, pre-defined detector regions mimic the output of conventional neural networks. The class with the highest sum of light intensity within its corresponding detector region, determined by \texttt{argmax}, is selected as the prediction result~\cite{lin2018all}. Suppose the ground truth class is \( t \), the loss function \( \boldsymbol{\ell} \) using \textbf{MSELoss} is defined as:
\[
\boldsymbol{\ell} = \| \text{Softmax}(I) - t \|_2.
\]
The DONN system is designed to be differentiable and compatible with conventional automatic differentiation engines.

\section{Experiments}
\label{sec:results}

We evaluate the performance of HoloGraph on graph-structured citation network datasets Cora-ML and CiteSeer from~\cite{sen2008collective}, and a node classification dataset Amazon Photo from~\cite{shchur2018pitfalls}. 

The default DONN system is constructed with six diffractive layers and the system size is set as $200 \times 200$, i.e., $N=200$. The system is configured with the laser wavelength $\lambda$ at $532nm$. 
We maintain a uniform physical distance $z$ of $27.94cm$ among layers, between the first layer and the source, and from the final layer to the detector. The distinct detector regions, corresponding to the number of classes, are uniformly situated on the detector plane. The aggregated intensity of the detector regions equates will return a vector in \texttt{float32} type. The final prediction results are computed using \texttt{argmax}. 
All implementations are constructed using PyTorch v1.13.1, and results are conducted on Nvidia Quadro RTX 6000 GPU using Pytorch-Geometric (PyG) framework.

\subsection{Node classification}
\label{sec:default_set}

In line with prior studies on node classification using benchmark graphs~\cite{rong2019dropedge}, we randomly select \(1000\) nodes for the test set in all benchmarks, leaving the remaining nodes for training purposes. To align with the scale of the DONN system, we implement principal component analysis (PCA) to preprocess the node attributes and compress the feature dimensions. The valid neighborhoods for feature aggregation are selected by PPRGo as the top-\(k\) ranked neighbor nodes.


In our evaluations with Cora and CiteSeer datasets, we configure the DONN architecture by setting the number of node attributes (\textit{dim}) to \(100\) after PCA dimension reduction and choosing the top-\(5\) ($k$ = 5) nodes for feature aggregation. For the Amazon Photo dataset, \textit{dim} is set to \(70\) with top-\(5\) neighbors. The input information is encoded on the amplitude part of the light signal, while setting the phase part as zero. As shown in Figure \ref{fig:sys_overview}, the number of diffractive layers in the DONN system is specified as \(6\) with the first three layers as feature extraction layers, while the last three layers as prediction layers. Three optical skip channels are implemented in the DONN system: the first one is implemented between the input layer (layer 0) and the first prediction layer (layer 4), the second one is implemented between layer 0 and layer 5, and the third one connects layer 0 and layer 6.
\vspace{-3mm}
\begin{table}[!ht]
\caption{{Node classification accuracy (\%) results on three graph benchmarks.}}
\label{best_result}
\centering

\begin{tabular}{c|ccc}
\toprule
\bf{Dataset}     & Cora-ML & Citeseer & Amazon Photo \\
\hline
PCA~\cite{abdi2010principal}              & 79.4    & 70.9     & 90.6 \\
MLP             & 81.5    & 71.3     & 93.1 \\
PPRGo~\cite{bojchevski2020scaling}            & \textbf{87.2}    & 74.3     & \textbf{95.0} \\
GCN~\cite{kipf2016semi}              & 87.0    & 77.1     & 94.5 \\
DGNN~\cite{yan2022all}             & 86.5    & 74.4     & 94.0 \\
\hline
HoloGraph (Ours) & 85.3    & \textbf{77.6}     & {94.8} \\
\toprule
\end{tabular}
\vspace{-3mm}
\end{table}

Table~\ref{best_result} displays the test accuracies for node classification using HoloGraph on benchmark graphs. These results are compared against traditional electronic computing methods, including linear principal components analysis (PCA), nonlinear multi-layer perceptrons (MLPs), nonlinear PPRGo GNN~\cite{bojchevski2020scaling}, Graph Convolutional Networks (GCN)~\cite{kipf2016semi}, and optical GNNs, Diffractive GNN~\cite{yan2022all}. 
The nonlinear PPRGo GNN employs an aggregator that conducts a weighted sum of neighboring features based on personalized PageRank scores. Both the MLP and the feature transformation of PPRGo utilize fully connected neural networks with two hidden layers, configured to have the same number of learnable parameters. The GCN model incorporates a 2-layer GCNConv layer with 16 hidden units. 
In the case of DGNN, we take the system setting with the best accuracy performance reported in \cite{yan2022all} for comparison.

The results presented in Table 1 underscore several key observations: (i) Models that leverage the graph structure (i.e., PPRGo, GCN, DGNN, and HoloGraph) exhibit significantly better performance than those that neglect it (i.e., PCA and MLP); (ii) our free-space all-optical graph learning model shows competitive performance with digital graph learning models. This implies that the free-space light diffraction optical modules and optical skip mechanism in HoloGraph are comparable to electronic message passing mechanism; and (iii) we achieve a classification accuracy similar to DGNN, with a $3.2\%$ higher accuracy on the Citeseer dataset and $0.8\%$ higher accuracy on the Amazon Photo dataset. This suggests that our free-space optical system can achieve similar performance with integrated optical systems while provide a substantial benefit in ease of fabrication on a large scale.
The convergence curves regarding both accuracy performance and loss, and the confusion matrix for all three datasets with HoloGraph, are provided in Appendix~\ref{appendix:loss_curve}. 

\subsection{Design Space Exploration}
\label{sec:res_dse}

In this section, we will explore the design space for HoloGraph including (1) DONN architecture exploration, i.e., the optical skip channels between different and multiple layers; (2) Input processing parameters explorations including the number of neighborhoods $k$, and the number of features $d$ of each node after feature compression by PCA, which are provided in Appendix~\ref{appendix:dse_input}.



\begin{figure}[!ht]
    \centering
    \begin{subfigure}[!h]{0.5\linewidth}
        \centering
        \resizebox{1\linewidth}{!}{%
        \begin{tabular}{c|c|c|c|c}
        \toprule
         & Name & \begin{tabular}[c]{@{}c@{}}$1^{st}$\\ Channel\end{tabular} & \begin{tabular}[c]{@{}c@{}}$2^{nd}$\\ Channel\end{tabular}& \begin{tabular}[c]{@{}c@{}}$3^{rd}$\\ Channel\end{tabular} \\
        \hline
        \begin{tabular}[c]{@{}c@{}}Input for\\ Aggregation\end{tabular} 
         & setup 1 & 0$\rightarrow$2 & 0$\rightarrow$3 & n/a \\
        \hline
        \begin{tabular}[c]{@{}c@{}}Input for\\ Prediction\end{tabular} 
         & setup 2 & 0$\rightarrow$4 & 0$\rightarrow$5 & 0$\rightarrow$6 \\
        \hline
        \multirow{2}{*}{\begin{tabular}[c]{@{}c@{}}Aggregation and\\ Prediction\end{tabular}} 
         & setup 3 & 1$\rightarrow$4 & 2$\rightarrow$4 & n/a \\
         & setup 4 & 3$\rightarrow$5 & 3$\rightarrow$6 & n/a \\
        \hline
        \multirow{2}{*}{\begin{tabular}[c]{@{}c@{}}Combined for\\ Prediction\end{tabular}} 
         & setup 5 & 0$\rightarrow$4 & 1$\rightarrow$4 & 2$\rightarrow$4 \\
         & setup 6 & 0$\rightarrow$4 & 1$\rightarrow$5 & 2$\rightarrow$6 \\
        \toprule
        \end{tabular}}
        \subcaption{}
        \label{tbl:optical_skip_setup}
    \end{subfigure}
    \hfill
    \begin{subfigure}[!h]{0.48\linewidth}
        \centering
        \includegraphics[width=1.1\linewidth]{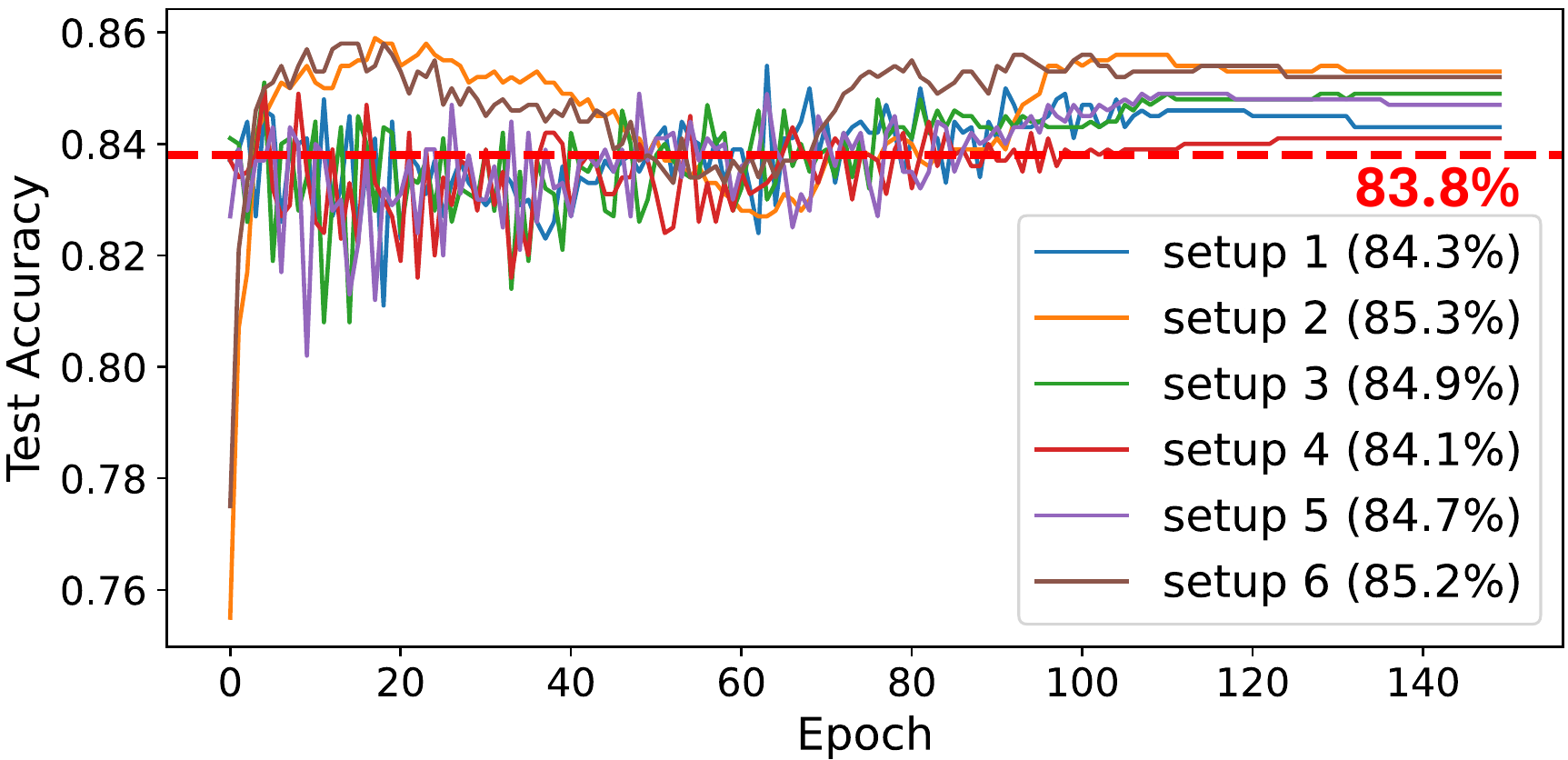}
        \subcaption{}
        \label{fig:skip_curve}
    \end{subfigure}

    \caption{(a) The system architecture exploration for setups of optical
skip connections in the proposed system. (b) Test accuracy w.r.t training epochs on Cora-ML under different optical skip connection setups in Table \ref{tbl:optical_skip_setup}.}
    \label{fig:skip_combined}
\end{figure}

\paragraph{DONN architecture exploration} 
This section explores the implementation of the optical skip channels in DONN systems. As shown in Figure~\ref{fig:sys_overview}, from the input layer as layer 0, the first three layers are feature embedding layers for feature aggregation (layer 1 — layer 3) and the last three layers are prediction layers for the classification predictions (layer 4 — layer 6). The system architecture is explored according to the system setups shown in Table~\ref{tbl:optical_skip_setup}. Specifically, we classify the setup explorations into three categories: (1) Input for feature aggregation (setup 1), where the original input is sent to all embedding layers, i.e., there are two optical skip channels connecting layer 0 to layer 2 and layer 3 separately. (2) Input for prediction (setup 2), where the original input is sent to all prediction layers, i.e., optical skip channels are connecting layer 0 to layer 4, layer 5, and layer 6 separately. (3) Aggregation and prediction, where the optical skip channels are implemented only between aggregation layers and prediction layers. We explore two message optical skip methods: first, all aggregations are connected to the first prediction layer (setup 3), i.e., the optical skip channels are connecting both layer 1 and layer 2 to layer 4; second, the aggregation is connected to all prediction layers (setup 4), i.e., the optical skip channels are connecting layer 3 to layer 5 and layers 6 separately. (4) Input and aggregation combined for prediction, where we combine both original input and intermediate aggregations to the predictions. We explore two connections: in the merged connection (setup 5), the optical skip channels are connecting layer 0, layer 1, and layer 2 all to layer 4; in the separate connection (setup 6), the optical skip channels are connecting layer 0 to layer 4, layer 1 to layer 5, and layer 2 to layer 6, separately. Note that the optical skip channels can be implemented between any arbitrary layers, we report total six setups from different categories for illustrations due to the page limit. 

\vspace{-3mm}

The system performance regarding all six setups (from setup 1 to setup 6) are shown in Figure~\ref{fig:skip_curve}, where the accuracy baseline from the system without optical skip connections is 83.8\% (red dashed line). From the results, we observe that: \textbf{(1)} the systems with optical skip connections always outperform the system without optical skip connections. \textbf{(2)} The original input information is important during the successive feature aggregation and predictions. As illustrated in Section \ref{sec:method}, after the light signal with the original input information is diffracted over the free space, the resulting diffraction pattern is significantly altered with inadvertently disseminated information beyond the designated computational space by the mathematical modeling for the system training. As a result, involving the skip channel connecting the original input to diffractive layers will restore and compensate the information loss during the propagation and results in better accuracy performance. \textbf{(3)} The separate skip connections (setup 6), where each previous layer connects to a different successive layer, performs better than the merged skip connections (setup 5), where the results from all previous layers are merged and connected the the same successive layer as the information after merged skip connections can be shuffled and confuse the DONN system, degrading the system performance. \textbf{(4)} Notably, there is a significant accuracy drop between epoch 20 to epoch 100 for setup 6 and setup 2. However, as shown in Figure \ref{fig:cvg_curve}(b), the loss curve during training with setup 6 remains stable. Thus, even through there is an accuracy drop, the system robustness is increasing during the training, indicating the necessity of training efforts regarding the system. 

\subsection{Hardware Performance Comparison}

Compared with conventional GNNs models on digital platforms, the current optical-devices-deployed HoloGraph DONN systems result in slightly accuracy performance degradation while feature with significantly improved energy efficiency. 
As shown in Table \ref{tbl:hw_comparison}, we evaluate the energy efficiency between GCN (the most energy efficient on the conventional side) and HoloGraph. We deploy the conventional GCN on Nvidia GPU 4090 Ti, Intel Xeon 6230 20x CPU, and estimate the energy efficiency using the SLMs-based hardware setup proposed in \cite{zhou2021large,li2023lightridge}. Regarding CPU/GPU results, we measure the consumed power using adaptive measurement, which only measures the power consumed for the GCN workload. With the advantage of free-space DONN system features with its high energy efficiency, we have achieved 3 -- 4 orders of magnitude improvements over GPU and CPU. Note that the HoloGraph energy efficiency can be further improved in a monolithic setting over the bulky passive prototype. 
Additionally, our HoloGraph architecture provides superior scalability and cost efficiency when compared with the existing integrated optoelectronic solution in \cite{yan2022all} as discussed in Section \ref{sec:compare_dgnn}.

\vspace{-3mm}
\begin{table}[!ht]
\caption{Energy efficiency comparison between CPU, GPU, and HoloGraph in frame/second/Joule. *One frame refers to one whole graph inference.}
\label{tbl:hw_comparison}
\centering
\begin{tabular}{c|ccc}
\toprule
\bf{Dataset} & \bf{4090 Ti} & \bf{Intel Xeon} & \bf{HoloGraph}  \\
\hline
Cora         & 1.004 (735$\times$)      & 0.031 (2.4e4$\times$)      & 738.552        \\
CiteSeer     & 0.637 (942$\times$)       & 0.063 (9.5e3$\times$)      & 601.142        \\
Amazon Photo & 0.093 (2.8e3$\times$)        & 0.020 (1.3e3$\times$)   & 261.438           \\
\toprule
\end{tabular}
\end{table}
\vspace{-5mm}

\section{Limitations and Discussion} \label{sec:dicussion}
\vspace{-2mm}
Our proposed HoloGraph architecture achieves competitive performance on graph learning tasks while offering significantly improved energy efficiency compared to existing graph learning solutions. By leveraging physics-based processing and optical skip channels between feature embedding layers and prediction layers in the DONN system, HoloGraph enriches the feature space for graph data. 
However, the more complex DONN architecture in HoloGraph introduces challenges in terms of stability, scalability, and generalizability.

First, the stability of DONN systems can be improved by more advanced fabrication technologies and monolithic system setups. For example, we can use the optoelectronic solution in \cite{zhou2021large} to implement diffractive layers by integrating digital control into the optical processing process for better system stability. Additionally, the metasurface-enabled on-chip integration of DONN systems in \cite{luo2022metasurface} offers a compact and stable DONN system at the chip scale. 


Second, the scalability of DONN systems is mainly limited by the optical energy loss incurred during free-space light propagation, particularly as the system architecture becomes more complex.  There are potential ways to improve the system scalability: (1) Increasing the system size, which allows the computational capacity to scale proportionally to the square of the system size; (2) Parallelizing computation by introducing additional optical channels via beam splitters; (3) Integrating digital control with optical processing to compensate for energy attenuation during propagation, such as the optoelectronic fused diffractive unit shown in \cite{zhou2021large}. 



Third, beyond static node classification tasks, our all-optical graph learning architecture offers distinct advantages for more energy-intensive and memory-demanding applications, such as dynamic graph learning~\cite{gurevin2023exploiting}. 
To enable DONN systems to process dynamic graphs, it is essential to encode temporal or structural variations effectively. 
As shown in Figure~\ref{fig:input_encode}, which has two dimensions for information encoding on the light signal, i.e., amplitude and phase. 
Node feature information is encoded on the amplitude dimension, while temporal or structural dynamics are encoded on the phase dimension. This dual-channel encoding allows the optical system to capture and process dynamic changes efficiently, aligning well with the computational requirements of dynamic graph tasks.



\section{Conclusion}\label{sec:conclusion}
\vspace{-2mm}
This paper proposes HoloGraph, the first monolithic free-space all-optical graph learning system, specifically designed to address the complexities of graph-structured tasks which have remained challenging for existing DONNs and similar physics-based neural networks. HoloGraph introduces a domain-specific message-passing mechanism with integrated optical skip channels, enabling light-speed optical message passing over graph structures. Through experiments on standard graph learning datasets, HoloGraph has demonstrated competitive and even superior classification performance compared to conventional GNNs. 
Moreover, our hardware deployment evaluations demonstrate significant energy efficiency improvements over the conventional GNNs on CPU/GPU platforms. 
Our ablation studies further validate the effectiveness of this novel architecture and the algorithmic methods introduced, underscoring HoloGraph's potential as a transformative solution in optical AI for graph learning. 

\newpage


\newpage
\bibliographystyle{plainnat}
\bibliography{ref}

@article{liu2023broadband,
  title={Broadband, Low-Crosstalk, and Massive-Channels OAM Modes De/Multiplexing Based on Optical Diffraction Neural Network},
  author={Liu, Zhibing and Gao, Shecheng and Lai, Zhaoyu and Li, Yuru and Ao, Zhaohuan and Li, Jinpei and Tu, Jiajing and Wu, Yanxiong and Liu, Weiping and Li, Zhaohui},
  journal={Laser \& Photonics Reviews},
  volume={17},
  number={4},
  pages={2200536},
  year={2023},
  publisher={Wiley Online Library}
}

@article{qian2020performing,
  title={Performing optical logic operations by a diffractive neural network},
  author={Qian, Chao and Lin, Xiao and Lin, Xiaobin and Xu, Jian and Sun, Yang and Li, Erping and Zhang, Baile and Chen, Hongsheng},
  journal={Light: Science \& Applications},
  volume={9},
  number={1},
  pages={59},
  year={2020},
  publisher={Nature Publishing Group UK London}
}

@inproceedings{li2023lightridge,
  title={LightRidge: An End-to-end Agile Design Framework for Diffractive Optical Neural Networks},
  author={Li, Yingjie and Chen, Ruiyang and Lou, Minhan and Sensale-Rodriguez, Berardi and Gao, Weilu and Yu, Cunxi},
  booktitle={Proceedings of the 28th ACM International Conference on Architectural Support for Programming Languages and Operating Systems, Volume 4},
  pages={202--218},
  year={2023}
}

@article{lin2018all,
  title={All-optical machine learning using diffractive deep neural networks},
  author={Lin, Xing and \textit{et al.}},
  journal={Science},
  volume={361},
  number={6406},
  pages={1004--1008},
  year={2018},
  publisher={American Association for the Advancement of Science}
}

@article{gilmer2020message,
  title={Message passing neural networks},
  author={Gilmer, Justin and Schoenholz, Samuel S and Riley, Patrick F and Vinyals, Oriol and Dahl, George E},
  journal={Machine learning meets quantum physics},
  pages={199--214},
  year={2020},
  publisher={Springer}
}

@article{zhou2021large,
  title={Large-scale neuromorphic optoelectronic computing with a reconfigurable diffractive processing unit},
  author={Zhou, Tiankuang and \textit{et al.}},
  journal={Nature Photonics},
  volume={15},
  number={5},
  pages={367--373},
  year={2021},
  publisher={Nature Publishing Group}
}

@article{chen2022physics,
  title={Physics-aware Complex-valued Adversarial Machine Learning in Reconfigurable Diffractive All-optical Neural Network},
  author={Chen, Ruiyang and \textit{et al.}},
  journal={arXiv preprint arXiv:2203.06055},
  year={2022}
}

@inproceedings{peng2021binary,
  title={Binary complex neural network acceleration on fpga},
  author={Peng, Hongwu and \textit{et al.}},
  booktitle={IEEE ASAP},
  pages={85--92},
  year={2021}
}

@article{yan2022all,
  title={All-optical graph representation learning using integrated diffractive photonic computing units},
  author={Yan, Tao and \textit{et al.}},
  journal={Science Advances},
  volume={8},
  number={24},
  pages={eabn7630},
  year={2022},
  publisher={American Association for the Advancement of Science}
}

@article{duvenaud2015convolutional,
  title={Convolutional networks on graphs for learning molecular fingerprints},
  author={Duvenaud, David K and \textit{et al.}},
  journal={Advances in neural information processing systems},
  volume={28},
  year={2015}
}

@inproceedings{hu2013matched,
  title={Matched signal detection on graphs: Theory and application to brain network classification},
  author={Hu, Chenhui and \textit{et al.}},
  booktitle={Information Processing in Medical Imaging: 23rd International Conference, IPMI 2013, Asilomar, CA, USA, June 28--July 3, 2013. Proceedings 23},
  pages={1--12},
  year={2013},
  organization={Springer}
}

@article{abdi2010principal,
  title={Principal component analysis},
  author={Abdi, Herv{\'e} and Williams, Lynne J},
  journal={Wiley interdisciplinary reviews: computational statistics},
  volume={2},
  number={4},
  pages={433--459},
  year={2010},
  publisher={Wiley Online Library}
}

@inproceedings{gilmer2017neural,
  title={Neural message passing for quantum chemistry},
  author={Gilmer, Justin and \textit{et al.}},
  booktitle={International conference on machine learning},
  pages={1263--1272},
  year={2017},
  organization={PMLR}
}

@article{velickovic2017graph,
  title={Graph attention networks},
  author={Velickovic, Petar and \textit{et al.}},
  journal={stat},
  volume={1050},
  number={20},
  pages={10--48550},
  year={2017}
}

@inproceedings{he2016deep,
  title={Deep residual learning for image recognition},
  author={He, Kaiming and \textit{et al.}},
  booktitle={CVPR},
  pages={770--778},
  year={2016}
}

@article{mengu2019analysis,
  title={Analysis of diffractive optical neural networks and their integration with electronic neural networks},
  author={Mengu, Deniz and \textit{et al.}},
  journal={JSTQE},
  volume={26},
  number={1},
  pages={1--14},
  year={2019},
  publisher={IEEE}
}

@article{mengu2020scale,
  title={Scale-, shift-, and rotation-invariant diffractive optical networks},
  author={Mengu, Deniz and \textit{et al.}},
  volume={8},
  number={1},
  pages={324--334},
  journal={ACS photonics},
  year={2020}
}

@article{shen2017deep,
  title={Deep learning with coherent nanophotonic circuits},
  author={Shen, Yichen and \textit{et al.}},
  journal={Nature photonics},
  volume={11},
  number={7},
  pages={441--446},
  year={2017},
  publisher={Nature Publishing Group}
}

@article{li2022physics,
  title={Physics-aware Differentiable Discrete Codesign for Diffractive Optical Neural Networks},
  author={Li, Yingjie and \textit{et al.}},
  journal={ICCAD},
  year={2022}
}

@article{tobin1997introduction,
  title={Introduction to Fourier Optics},
  author={Tobin, Mary S},
  journal={American Scientist},
  volume={85},
  number={6},
  pages={581--584},
  year={1997},
  publisher={Sigma Xi, The Scientific Research Society}
}

@inproceedings{bojchevski2020scaling,
  title={Scaling graph neural networks with approximate pagerank},
  author={Bojchevski, Aleksandar and Gasteiger, Johannes and Perozzi, Bryan and Kapoor, Amol and Blais, Martin and R{\'o}zemberczki, Benedek and Lukasik, Michal and G{\"u}nnemann, Stephan},
  booktitle={Proceedings of the 26th ACM SIGKDD International Conference on Knowledge Discovery \& Data Mining},
  pages={2464--2473},
  year={2020}
}

@article{rong2019dropedge,
  title={Dropedge: Towards deep graph convolutional networks on node classification},
  author={Rong, Yu and Huang, Wenbing and Xu, Tingyang and Huang, Junzhou},
  journal={International Conference on Learning Representations},
  year={2019}
}

@article{shchur2018pitfalls,
  title={Pitfalls of graph neural network evaluation. arXiv 2018},
  author={Shchur, Oleksandr and Mumme, Maximilian and Bojchevski, Aleksandar and G{\"u}nnemann, Stephan},
  journal={arXiv preprint arXiv:1811.05868},
  year={2018}
}

@article{kipf2016semi,
  title={Semi-Supervised Classification with Graph Convolutional Networks},
  author={Kipf, Thomas N and Welling, Max},
  journal={arXiv preprint arXiv:1609.02907},
  year={2016}
}

@article{sen2008collective,
  title={Collective classification in network data},
  author={Sen, Prithviraj and Namata, Galileo and Bilgic, Mustafa and Getoor, Lise and Galligher, Brian and Eliassi-Rad, Tina},
  journal={AI magazine},
  volume={29},
  number={3},
  pages={93--93},
  year={2008}
}

@article{luo2022metasurface,
  title={Metasurface-enabled on-chip multiplexed diffractive neural networks in the visible},
  author={Luo, Xuhao and Hu, Yueqiang and Ou, Xiangnian and Li, Xin and Lai, Jiajie and Liu, Na and Cheng, Xinbin and Pan, Anlian and Duan, Huigao},
  journal={Light: Science \& Applications},
  volume={11},
  number={1},
  pages={158},
  year={2022},
  publisher={Nature Publishing Group UK London}
}

@article{gurevin2023exploiting,
  title={Exploiting Intrinsic Redundancies in Dynamic Graph Neural Networks for Processing Efficiency},
  author={Gurevin, Deniz and Ding, Caiwen and Khan, Omer},
  journal={IEEE Computer Architecture Letters},
  volume={23},
  number={2},
  pages={170--174},
  year={2023},
  publisher={IEEE}
}

\newpage
\appendix
\section{Training curves for results in Node Classification}
\label{appendix:loss_curve}
\begin{figure}[!ht]
  \centering
   \subcaptionbox{Accuracy convergence plot.}
  {\includegraphics[width=0.4\columnwidth]{  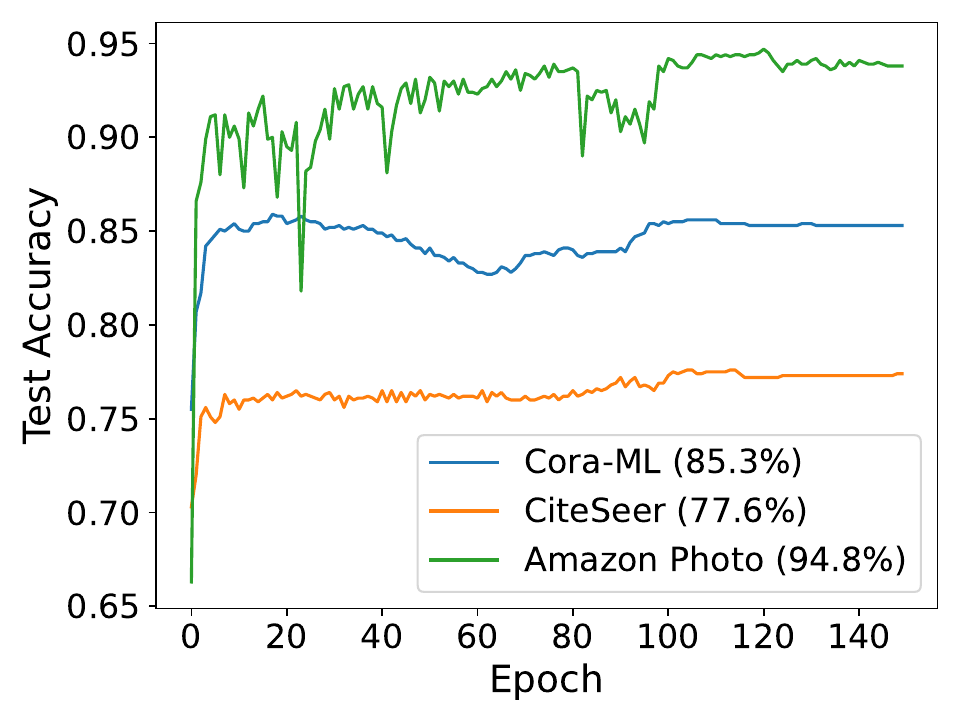}}
    \subcaptionbox{Loss convergence plot.}  {\includegraphics[width=0.4\columnwidth]{  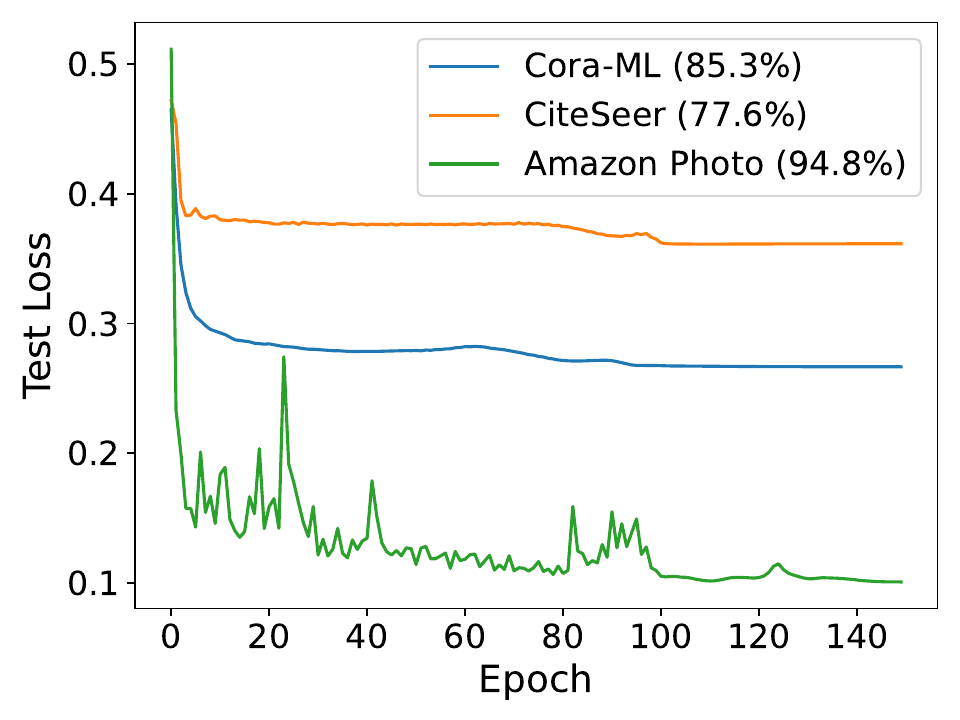}}
\caption{Accuracy and Loss convergence plots of HoloGraph on Cora, CiteSeer and Amazon Photo benchmarks.}
\label{fig:cvg_curve}
\end{figure}

\begin{figure}[!ht]
  \centering
   \subcaptionbox{Cora}
  {\includegraphics[width=0.32\textwidth]{  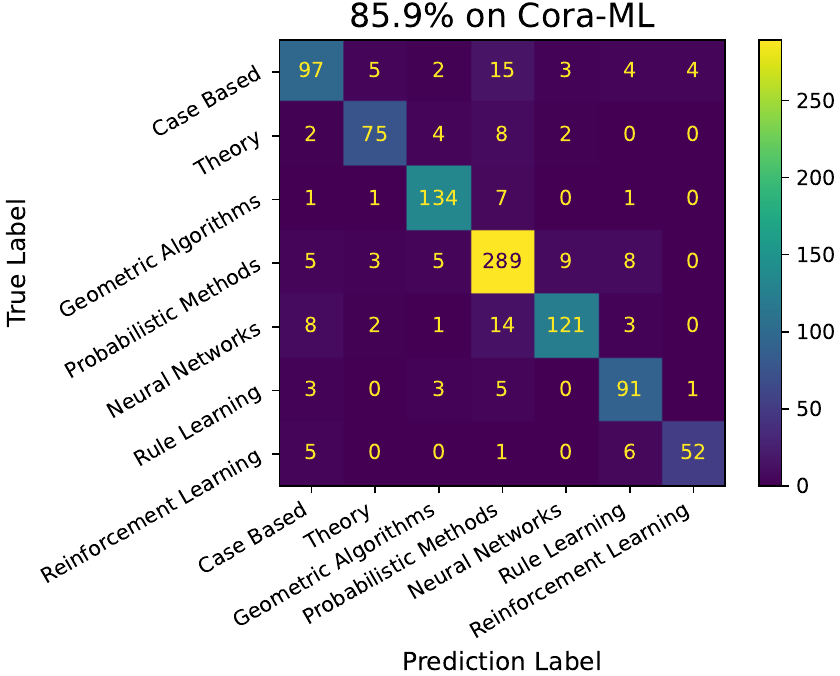}}
    \subcaptionbox{Citeseer}
  {\includegraphics[width=0.32\textwidth]{  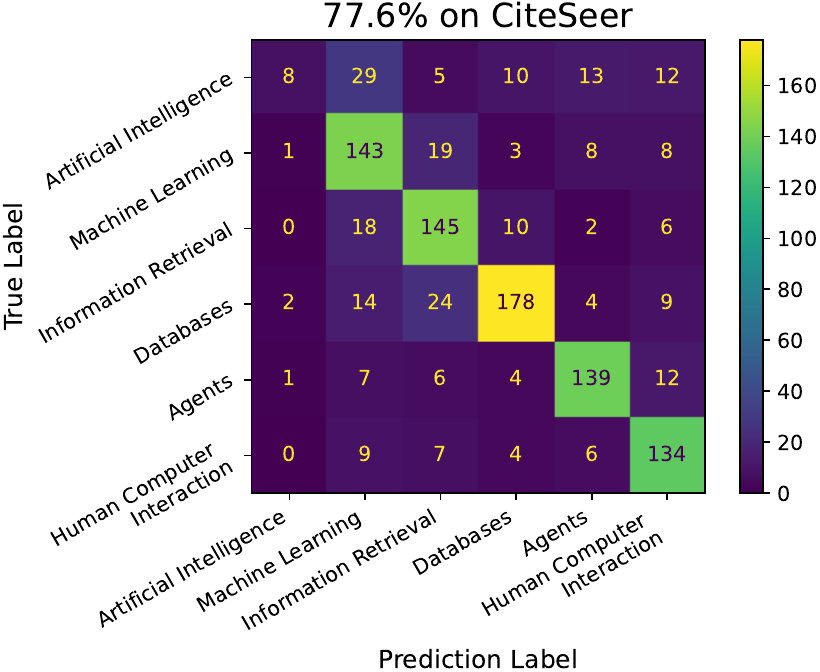}}
    \subcaptionbox{Amazon Photo}
  {\includegraphics[width=0.32\textwidth]{  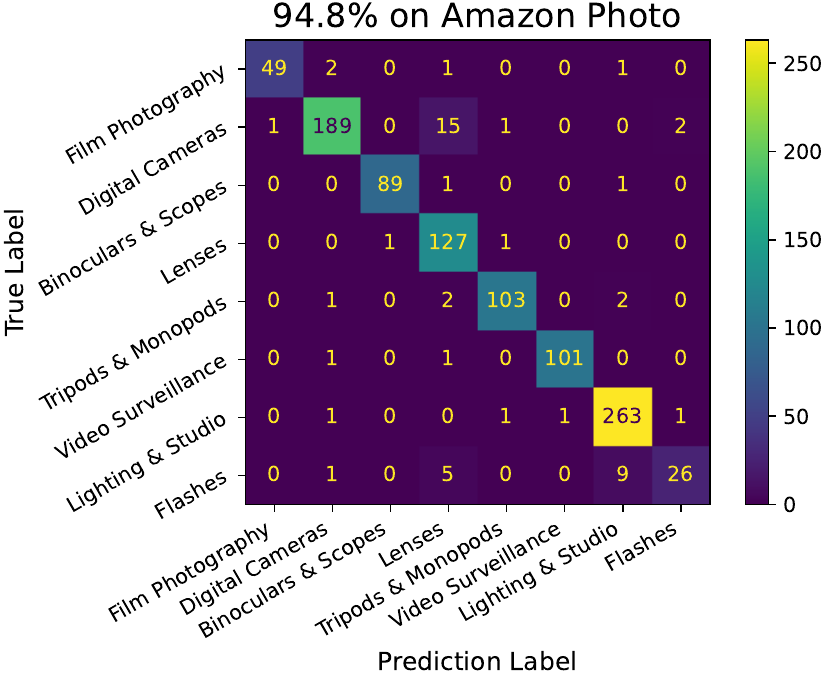}}
\caption{Confusion Matrix of HoloGraph classification result on three datasets.}
\label{fig:cfm}
\end{figure}

The convergence curves regarding both accuracy performance and loss are provided in Figure~\ref{fig:cvg_curve}. For smaller datasets like Cora and CiteSeer, the performance converges more efficiently and more stably while for larger dataset like Amazon Photos, it requires more training efforts to convergence. 
Additionally, the confusion matrix for all three datasets with HoloGraph is shown in Figure~\ref{fig:cfm}.

\section{Input processing parameters exploration}\label{appendix:dse_input}
We explore the parameters for input data processing in this section targeting the number of neighborhoods $k$, the number of features $d$, and the encoding with score from the PageRank for neighbor selections on the phase of the input wavefunction. The system size fixed at $200$ and the exploration is conducted on the Cora dataset.

\begin{figure}
     \centering
     \begin{subfigure}{0.45\textwidth}
         \centering
         \includegraphics[width=1\columnwidth]{  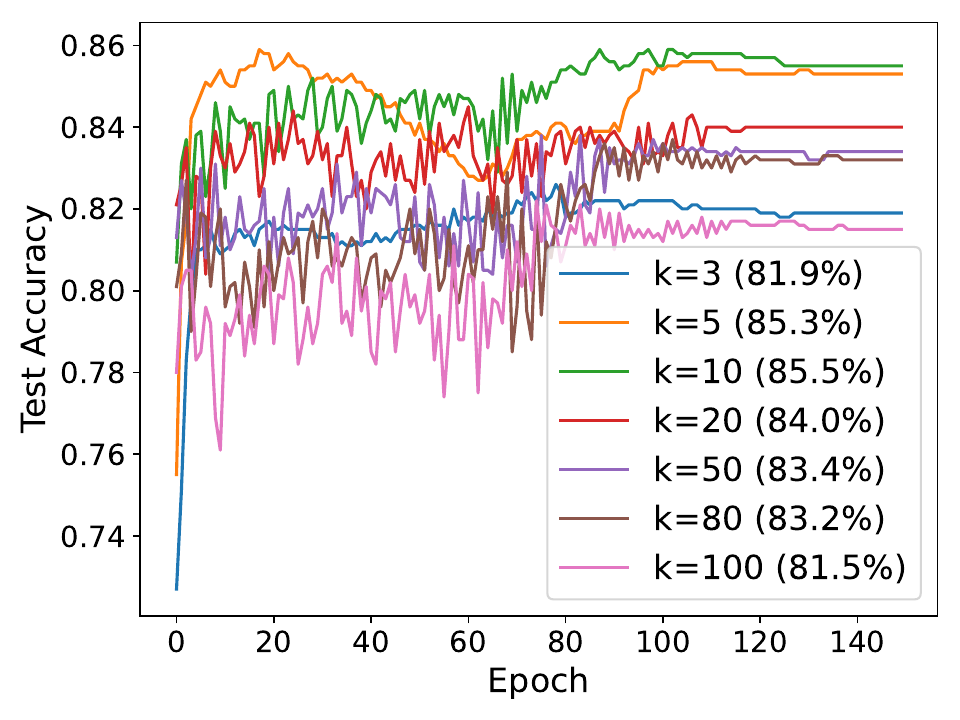}
         \caption{Test accuracy w.r.t training epochs under different No. of neighbors (d=100).}
         \label{fig:ablation_k}
     \end{subfigure}
     \hfill
    \begin{subfigure}{0.45\textwidth}
         \centering
        \includegraphics[width=1\linewidth]{  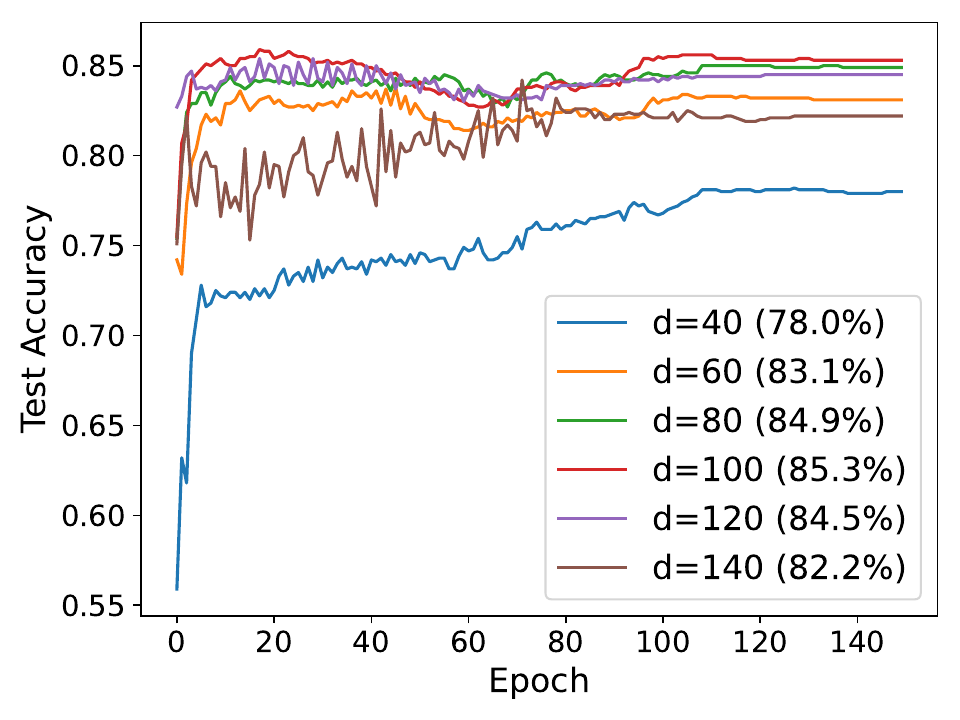}
         \caption{Test accuracy w.r.t training epochs under different No. of feature dimensions (k=5).}
         \label{fig:ablation_d}
     \end{subfigure}

        \caption{{Ablation study of HoloGraph w.r.t (a) top-$k$ neighborhoods and (b) $d$-dimensional input node attributes on Cora-ML for node classification.}}
        \label{fig:ablation_d_k}
\end{figure}
\paragraph{Study 1: Top-$k$ neighborhoods selection} -- For the exploration regarding $k$, we increase the number of neighborhoods for feature aggregation $k$ from 3 to 100 while keeping other parameters as default settings described in Section ~\ref{sec:default_set}.  The test accuracy curves of the setups are shown in Figure~\ref{fig:ablation_k}. When less neighborhoods are implemented for the feature aggregation, the system performance degrades significantly due to the lack of input information while the convergence is stable. However, when too many neighborhoods are taken, the accuracy performance also degrades because of the redundant and misleading information and the accuracy curves become significantly unstable during the training process, especially for the first 100 epochs. The setup with $k=10$ shows the best performance regarding convergence and accuracy. 

\paragraph{Study 2: Feature dimension $d$} -- For the exploration with $d$, we increase $d$ from 40 to 140 while other parameters are set as default shown in Section~\ref{sec:default_set}. The results are shown in Figure~\ref{fig:ablation_d}. When the original features are compressed to less dimensions (d $\leq$ 100), the system shows degraded accuracy performance due to the lack of accurate information. However, with higher dimensional features, the performance degrades due to information redundancy and model overfitting. The training process is more smooth for setups with less feature dimensions than that with higher feature dimensions. We pick the best setup with $d=100$ as our default setting. 

\begin{figure}[!htb]
     \centering
     \begin{subfigure}[b]{0.4\textwidth}
         \centering
         \includegraphics[width=0.98\columnwidth]{  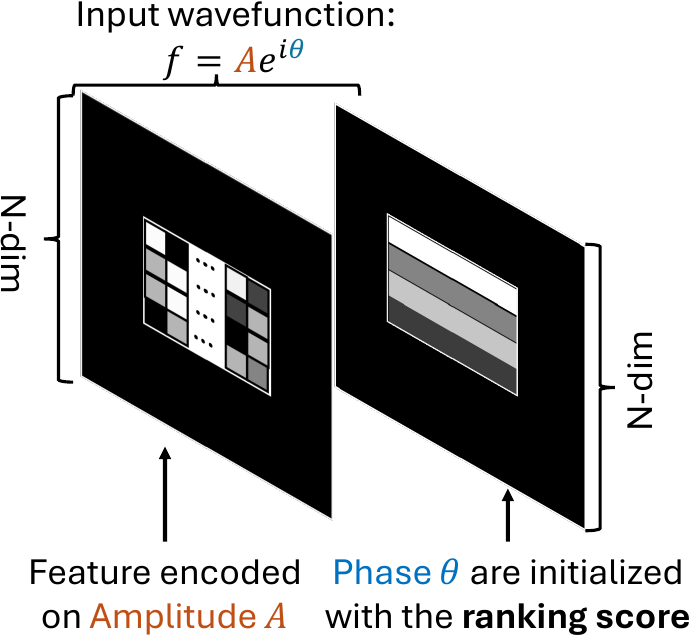}
         \caption{The input with neighbors' ranking score encoded on phase.}
         \label{fig:score_input}
     \end{subfigure}
    \begin{subfigure}[b]{0.4\textwidth}
         \centering
        \includegraphics[width=1\linewidth]{  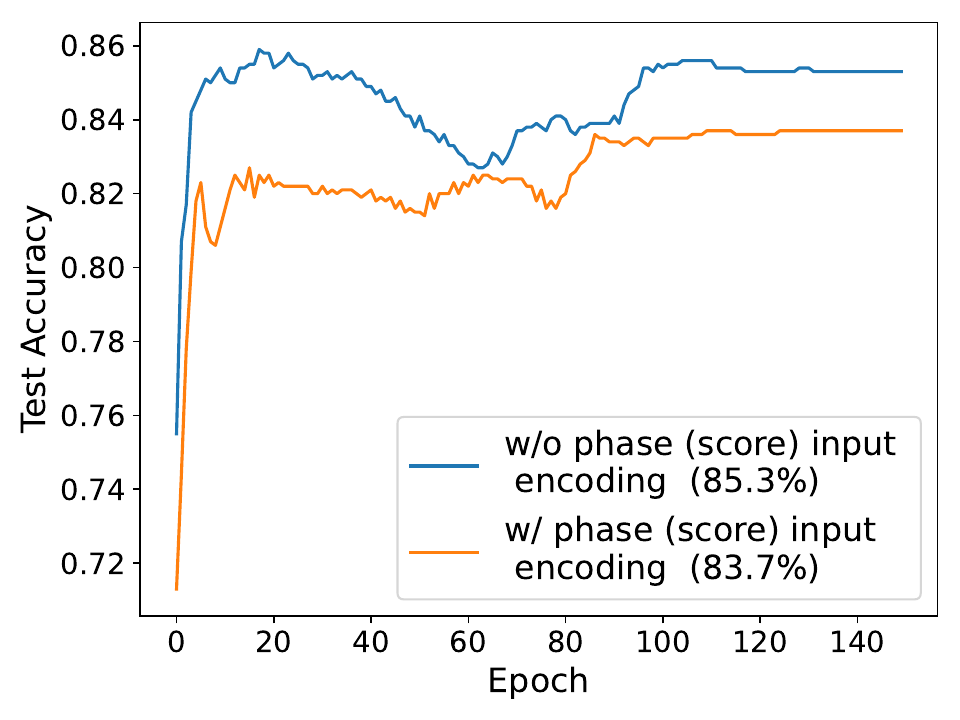}
         \caption{Accuracy convergence plots w/ and w/o score encoded on the input phase (k=5, d=100).}
         \label{fig:acc_score}
     \end{subfigure}

        \caption{{The exploration of the input encoding w and w/o the ranking score from PPGRo for each neighbor encoded on the phase of the light wavefunction.}}
        \label{fig:ablation_score}

\end{figure}

\paragraph{Study 3: w and w/o neighbor ranking score encoded with the input phase} -- For the default input encoding, only the features are encoded with the amplitude of the light wavefunction while the phase is set to zeros as shown in Figure~\ref{fig:input_encode}. In this section, we further include the score for each neighborhood resulting from the personalized PageRank in the input as shown in Figure~\ref{fig:score_input}. Specifically, besides the features encoded with the amplitude, the scores are encoded with the phase of the input wavefunction. The results are shown in Figure~\ref{fig:acc_score}. Even though the encoding with the score information is more informative, it results in 1.6\% accuracy degradation compared to the default encoding. This can result from the information alteration due to the light diffraction. When the input is equipped with phase information, the input information at the first diffractive layer is compressed and shuffled. 

\section{Comparison with DGNN}
\label{sec:compare_dgnn}
Compared to the integrated DGNN hardware system \cite{yan2022all}, the free-space HoloGraph framework offers superior scalability and cost efficiency. 
In the DGNN system, optical message passing is implemented with integrated DPUs, and node features are aggregated using Y-couplers. To avoid waveguide crossings, the DGNN employs a two-dimensional optical message representation. When performing graph learning with $k$ neighbors, the cost of the Y-couplers scales quadratically with $k^2$, while the cost of DPUs scales linearly with $k$ for a single aggregation head. Additionally, both Y-couplers and DPUs incur costs that are proportional to the number of node attributes $d$. Consequently, the total cost of the DGNN hardware system is proportional to $d \times (k + k^2)$. In contrast, HoloGraph leverages the high resolution of free-space optical devices, typically ranging from hundreds to thousands, which enables easy scalability. As the graph size increases, the system can be scaled by enlarging the mask size, laser beam diameter, and the active detection area on the detector, where the cost scales linearly with $k\times d$.



\end{document}